# Negative Databases for Biometric Data

Julien Bringer[1] and Hervé Chabanne[1,2]
[1] Sagem Sécurité, France.
[2] Télécom ParisTech, France.

November 11, 2018

**Abstract**

Negative databases – negative representations of a set of data – have been introduced in 2004 to protect the data they contain. Today, no solution is known to constitute biometric negative databases. This is surprising as biometric applications are very demanding of such protection for privacy reasons. The main difficulty comes from the fact that biometric captures of the same trait give different results and comparisons of the stored reference with the fresh captured biometric data has to take into account this variability. In this paper, we give a first answer to this problem by exhibiting a way to create and exploit biometric negative databases.

**Keywords.** Biometric data, Negative database, Privacy.

## 1 Introduction

Biometric data must be protected in order to prevent someone to be able to track back the users of a biometric system. Recently, they have been a lot of researches on their storage in a way which is renewable and which does not leak information. Typically, biometric data are quantized and encrypted. This encryption must still permit the matching of the underlying biometric data without decrypting them. On one hand, some very simple techniques of encryption, known as secure sketches have been suggested [10, 19] but their resistance seems doubtful in practice [2, 20]. On the other hand, secure sketches can be combined with homomorphic encryption but in this case, the performances of the computations are penalized [1, 5, 21].

In this paper, we have a different approach following the one of the negative databases [12]. In a negative database, instead of having the elements of a database $\mathcal{DB}$, we consider the complementary $\overline{\mathcal{DB}}$ of these elements. This means that instead of checking whether $b \in \mathcal{DB}$, we have to equivalently verify that $b \notin \overline{\mathcal{DB}}$. The representation of negative databases is made possible thanks to a wild-card symbol $*$ which stands for all the values; for instance, as we are here going to work with binary vectors, a $*$ for a bit means either the value 0 or 1. Negative databases got very interesting properties. Firstly, for a given database $\mathcal{DB}$, different negative databases $\overline{\mathcal{DB}}$ can be established. Moreover, starting from $\overline{\mathcal{DB}}$, it is hard to retrieve $\mathcal{DB}$. Finally, relational algebra also exists for negative databases. It should be noted that our approach is different from the previous one which treats biometric data individually while we are here considering a database as a whole.

The difficulty we encounter is that each capture of the same biometric data gives a different value. We here consider binarized biometric data, i.e. $b$ stands for a binary vector representing a biometric trait. A new capture of this biometric trait will give a binary vector with numerous coordinates differing from the ones of $b$. Our approach follows the one described in for identifying people thanks to their iris [16]. We here how show to handle this problem.

We begin in Section 2 by recalling some works on the binarization of biometric data. We state the properties these binarized data have to fulfill for the rest of our work. In Section 3, we give a short introduction to negative databases. Section 4 constitutes the core of our proposal and explains how to create and use biometric negative databases. We give an example to gauge the efficiency of our solution when applied to a real-life case. And finally, Section 6 concludes.

We conclude this introduction by recalling some basic facts about biometric data. A reader, familiar with this topic, can skip it to go directly to Section 2.

## 1.1 Biometric Systems: 101

Biometric recognition techniques can lead to quite different applications than those which are possible when you are dealing with, for instance, passwords. A major difference comes from the fact that your biometric data enable to identify yourself among a large set of people during your whole lifetime. This characteristic is reinforced by the fact that biometric data are non-transferable to someone else. On one hand, a positive aspect is that this can been seen as a very natural and easy-to-deploy way of identifying populations (think at the census of the citizens of a country). On the other hand, biometric data must be protected to respect their privacy. Today, AFIS (Automated Fingerprint Identification System) are present in many countries worldwide for police or civil applications. These AFIS can gather together the biometric data of millions of users. Usually, for measuring their performances, we evaluate their accuracy in terms of FAR (False Acceptance Rate: the probability that the system rejects a genuine user) and FRR (False Reject Rate: the probability that the system accepts an impostor). These 2 rates FAR and FRR cannot be reduced both at the same time and some compromise must be found according your wish to favour security or comfort. In this paper, we are looking at biometric systems of smaller scale. Typically, we are considering biometric readers mainly used for restricting the access control to a building or a room. The need for storage of biometric data is then limited to less than one hundred records.

## 2 Binarization of Biometric Data

Let $\beta$ designate the biometric trait of an user. Let $b \leftarrow \beta$ indicate that the value $b$ has been captured by a sensor.

We here make the hypothesis that the biometric $b$ are quantized and can be presented as binary vectors of length $n$, $b \in \{0,1\}^n$ in such a way that:

**Condition 1**   1. *Two different captures $b, b'$ from the same user $\mathcal{U}$ are with high probability at a Hamming distance $d(b, b') \leq \lambda_{min}$.*

2. *Captures $b_1, b_2$ of different users $\mathcal{U}_1, \mathcal{U}_2$ are at a Hamming distance $d(b_1, b_2) > \lambda_{max}$.*

The origin of Condition 1 comes from iris recognition technology [8] where the concept of iriscodes has been introduced. Iriscodes are binary vectors of length $n = 2048$ where the features of iris are represented. They are compared by their Hamming distance. Following this, various attempts have been made for applying this kind of representations to fingerprints [3, 18] or faces as in [6].

Condition 1 has been exploited in [16] for an efficient search algorithm over a large database. With a high probability the vectors we are comparing will have many small portions of their coordinates equal whenever they come from the same user as they are close for the Hamming distance. Let $\mathcal{DB} = \{b_1, \ldots, b_N\}$. [16] uses $h_1, \ldots, h_{128}$, projections of binary vectors (here iriscodes) over a part of their coordinates. They consider that a freshly captured iris $b$ can match an element stored in $\mathcal{DB}$ whenever:

$$h_{j_1}(b) = h_{j_1}(b_k),\ h_{j_2}(b) = h_{j_2}(b_k),\ h_{j_3}(b) = h_{j_3}(b_k) \tag{1}$$

for $1 \leq j_1 < j_2 < j_3 \leq 128$ and $1 \leq k \leq n$.

We here skip the details. This way of proceeding will serve us as the basis of our work on biometric negative databases which is described in Section 4. We recall a definition that formalizes this notion and which has been introduced for approximate nearest neighbor search:

**Definition 1 (Locality-Sensitive Hashing, [17])** *Let $(E, d)$ be the Hamming space, $F$ be a set, $r_1, r_2 \in \mathbb{R}$ with $r_1 < r_2$, $p_1, p_2 \in [0, 1]$ with $p_1 > p_2$. Let $\mathcal{H} = \{h_1, \ldots, h_L\}$ be a family of functions $h_i : E \to F$.*

*The family $\mathcal{H}$ is $(r_1, r_2, p_1, p_2)$-LSH, if*

$$\forall x, x' \in E \begin{cases} Pr_{h \in \mathcal{H}}[h(x) = h(x') \,|\, d(x, x') < r_1] > p_1 \\ Pr_{h \in \mathcal{H}}[h(x) = h(x') \,|\, d(x, x') > r_2] < p_2 \end{cases}$$

In the sequel we will assume that the inequalities above are equalities, i.e. the first (resp. second) probability is equal to $p_1$ (resp. $p_2$). This holds for the hash constructions used by [16].

In practice, we apply Condition 1 to a family of Locality-Sensitive Hashing functions that are combined to obtain the following result:

**Proposition 1** *Let the family $\mathcal{H} = \{h_1, \ldots, h_L\}$ be a $(\lambda_{min}, \lambda_{max}, p_1, p_2)$-LSH family. Let m be the number of hash functions that we consider simultaneously, following the principle of Equation (1).*

*The probability not to output 'matching' for a genuine user – called* **False Reject Rate (FRR)** *– is approximately*

$$P_{fr} = \sum_{i=0}^{m-1} \binom{L}{i} p_1^i (1-p_1)^{L-i}$$

*and the probability to output 'matching' for an impostor – called* **False Accept Rate (FAR)** *– is approximately*

$$P_{fa} = \sum_{i=m}^{L} \binom{L}{i} p_2^i (1-p_2)^{L-i}.$$

The variable $m$ is called **order of the hash equalities** in the sequel.

*Proof.* Let $b_1$, $b_2$ be two different biometric captures (possibly coming from the same user). They are considered as a matching pair as soon there is at least $m$ functions $h_{i_1}, \ldots, h_{i_m}$ from $\mathcal{H}$ such that

$$h_{i_1}(b_1) = h_{i_1}(b_2), \ldots, h_{i_m}(b_1) = h_{i_m}(b_2).$$

As $\mathcal{H}$ is $(\lambda_{min}, \lambda_{max}, p_1, p_2)$-LSH, from Condition 1, we know that with a high probability $Pr_{h \in \mathcal{H}}[h(b_1) = h(b_2)] = p_1$ if $b_1, b_2$ come from the same user and $Pr_{h \in \mathcal{H}}[h(b_1) = h(b_2)] = p_2$ if they come from different users. □

## 3  Negative Databases

A negative database consists of the representation of the negative image of a given database. The goal is to represent all the elements not in the original database instead of storing explicitly the data itself. The main issue is to find a way to represent concisely the negative database without letting the possibility to retrieve easily the original records. Note that there exists an extended notion of negative database where almost all elements not in the original database are represented. In this section, we refer only to the first notion for which the definition – inspired by by Esponda et al. [12] and subsequent works – follows.

**Definition 2** *Let $\mathcal{DB}$ be a database containing N vectors. Let l denote the length of the binary vectors belonging to $\mathcal{DB}$. A negative database $\overline{\mathcal{DB}}$ of $\mathcal{DB}$ is a finite set of vectors such that*

- *there is an algorithm IsMember$_{\overline{\mathcal{DB}}}$ enabling to check efficiently whether one l bits string is a member of $\overline{\mathcal{DB}}$;*

- *$x \in \{0,1\}^l$ is an element represented by $\overline{\mathcal{DB}}$ if and only if $x \notin \mathcal{DB}$, i.e. that $\overline{\mathcal{DB}}$ represents $\{0,1\}^l - \mathcal{DB}$.*

If $l$ is large, then the number of elements represented by $\overline{\mathcal{DB}}$ could be huge. To achieve a compact representation, [14] introduced the wild-card symbol '*', with the classical role: a position set to * in a string represents both 0 and 1 values. This doing a negative database is composed of vectors of length $l$ with the alphabet $\{0, 1, *\}$. The associated $\mathsf{IsMember}_{\overline{\mathcal{DB}}}$ algorithm corresponds to the string matching algorithm; a element $x$ is a member if $\overline{\mathcal{DB}}$ contains a vector $y$ for which simultaneous binary valued positions of $x$ and $y$ should be equal:

$$\forall i \in \{1, \ldots, l\}, (x_i = y_i \text{ OR } x_i = * \text{ OR } y_i = *).$$

Several polynomial-time algorithms have been published for negative databases computation [7, 9, 13] with quite different techniques. We recall in Table 1 the algorithm introduced in [12] to compute a negative representation of a database thanks to the prefix method. $w_i$ will stand in the following to a prefix of length $i$ and $W_i$ will represent the set of vectors of length $i$ (a part of the $w_i$'s).

1. $i \leftarrow 0$
2. $W_i \leftarrow \{\}$
3. $W_{(i+1)} \leftarrow$ set of vectors of length $i+1$ with a prefix of length $i$ in $W_i$ which is not a prefix of an element of $\mathcal{DB}$
4. for all $x$ in $W_{i+1}$
5. output a vector $y$ with $x$ as a prefix and complete it with *'s up to length $l$
6. add to $\overline{\mathcal{DB}}$
7. $i \leftarrow i + 1$
8. $W_i \leftarrow$ set of prefixes of length $i$ in $\mathcal{DB}$
9. go back to step 3 while $i < l$

Table 1: Prefix algorithm for negative representation

As proved in [14], the prefix algorithm outputs a negative database in $lN$ time (where $N$ is the number of records of $\mathcal{DB}$) and the resulting negative database will have at most $lN$ entries (elements of $\{0, 1, *\}^l$). This leads for $\overline{\mathcal{DB}}$ to an overall size $(2 \times N \times l^2)$.

The prefix algorithm shows the feasibility of negative representation. As for security concerns, [14] demonstrated that the reconstruction of $\mathcal{DB}$ from such a negative database with alphabet $\{0, 1, *\}$ is NP-hard, based on reduction to 3-SAT problem. Concerning the way to construct hard instances which resist to SAT solvers, the known algorithms produce negative representations with different level of security, depending for example on the number of *'s used. See [9, 13] for solutions on how to generate such hard instances. One another interesting property is the capability to create several different randomized negative representations from the same $\mathcal{DB}$.

To obtain randomized representation, [13] suggests a non deterministic version of the prefix algorithm based on:

- the use of random permutation to include the wild-card symbol also at the beginning or in the middle of a vector;
- a random replacement of wild-card by both the values 0 and 1.

The resulting algorithm is designed to run in time $l^2 N^2$ and produces negative databases of size $2 \times N \times l^3$.

From a given negative database, it is even possible to transform it into a new negative representation which is the negative image of the same database, but for which the two negative databases are different and it is difficult to determine if they are equivalent. In [11], this operation is denoted Morph.

Together with the creation of a negative representation, specific operations on $\mathcal{DB}$ are translated in operations applied on $\overline{\mathcal{DB}}$. Inserting (resp. deleting) a string into (resp. from) $\overline{\mathcal{DB}}$ corresponds to remove (resp. insert) the corresponding binary strings from (resp. into) $\mathcal{DB}$. As for the creation algorithm, randomized operations are also available [12]. Then an alternative solution for creating randomized representation is to start with a randomized negative representation of the empty set and to delete the data corresponding to $\mathcal{DB}$.

As remarked by [13], using these operations many times may have the effect on the size of $\overline{\mathcal{DB}}$ to grow unreasonably. An algorithm, somehow to clean regularly the representation, is designed to control this bad effect (cf. Clean-Up algorithm [13]).

Moreover, more complex operations are designed in [15] in the field of relational algebra operators (equality, less-than, union, cartesian product, intersection, ...). Each operator on $\mathcal{DB}$ has its transposition as an negative operator on $\overline{\mathcal{DB}}$.

Note that several other techniques are analyzed in the literature. For instance [7] introduced a negative representation with an higher overall growth factor but for which the security relies on cryptographic hash properties.

## 4 Our Solution

We explain now how to manage biometric data via negative representation in order to protect the content of the enrollment database. The main motivation is authorization scenario, nonetheless our construction can also be used for authentication.

### 4.1 Biometric Negative Database

We recall that our database $\mathcal{DB} = \{b_1, \ldots, b_N\}$. Let $\mathcal{H} = \{h_1, \ldots, h_L\}$ be $(\lambda_{min}, \lambda_{max}, p_1, p_2)$-LSH family of functions as defined in Definition 1 Section 2. Let $m$ be the order of the hash equalities, i.e. the number of these functions we are using as in Prop. 1. We are now ready to define the biometric negative database $\overline{\mathcal{DB}}$.

For $1 \leq k \leq N$, let $\mathcal{DB}_{b_k}$ be the database made of the elements

$$j_1 || \ldots || j_m || h_{j_1}(b_k) || \ldots || h_{j_m}(b_k) \tag{2}$$

with $1 \leq j_1 < \ldots < j_m \leq L$ and where $||$ stands for the concatenation.

**Definition 3** *An $(\mathcal{H}, m, \mathcal{DB})$-biometric negative database $\overline{\mathcal{DB}}$ is a negative representation of the dataset $\mathcal{U} = \bigcup_{k=1}^{N} \mathcal{DB}_{b_k}$ which is made of all order $m$ hash chains obtained with respect to $\mathcal{DB}$ and the LSH family $\mathcal{H}$ (cf. Equation (2)).*

The algorithm to determine if one fresh capture $b'$ is close to one enrolled template within the original dataset $\mathcal{DB}$ via one of its biometric negative database $\overline{\mathcal{DB}}$ is explained by Table 2.

---

**Input:** fresh biometric template $b'$
1. result← OK
2. For all $1 \leq j_1 < \ldots < j_m \leq L$
3.    $z \leftarrow j_1 || \ldots || j_m || h_{j_1}(b') || \ldots || h_{j_m}(b')$
4.    If $\mathsf{IsMember}_{\overline{\mathcal{DB}}}(z) = \text{OK}$
5.      Then
6.        result← NOK
7.        Break
8.      Else Continue
9. End For
10. Output result

---

Table 2: Authorization Check for Biometric Negative Database

This construction enables us to achieve the following properties.

**Proposition 2** *Let $\overline{\mathcal{DB}}$ be an $(\mathcal{H}, m, \mathcal{DB})$-biometric negative database.*

1. *$\overline{\mathcal{DB}}$ is not a negative database of $\mathcal{DB}$.*

2. *$\overline{\mathcal{DB}}$ enables to check whether a fresh capture $b'$ is close to one element of $\mathcal{DB}$ by verifying whether all the derived hash chains following Eq. (2) are not in $\overline{\mathcal{DB}}$. The error rates of this membership checking operation are:*

    **False 'Not Member' Decision**
    $$P_{fr}(1 - P_{fa})^{N-1}$$

    **False 'Member' Decision**
    $$1 - (1 - P_{fa})^N.$$

*Proof.*

1. The first statement comes from the fact that $\overline{\mathcal{DB}}$ is a negative database of $\mathcal{U} = \bigcup_{k=1}^{N} \mathcal{DB}_{b_k}$, which is not equivalent to $\mathcal{DB}$, thanks to Proposition 1 (except in the case where $p_1, p_2$ are negligible, which corresponds to the situation where the hash functions are cryptographic ones with hard-to-find collisions; i.e. the case where almost only original data satisfy the order $m$ hash equalities).

2. The second statement is induced together by Proposition 1 and Definition 2. The algorithm $\mathsf{IsMember}_{\overline{\mathcal{DB}}}$ gives a way to decide whether data are in $\mathcal{U}$ from the negative test of presence in $\overline{\mathcal{DB}}$. Moreover, Proposition 1 implies that for a genuine user, the probability not to find an order $m$ hash chain following Eq. (2) in $\mathcal{U}$ is approximately

$$P_{fr}(1 - P_{fa})^{N-1}$$

(i.e. that you find neither any genuine equalities nor any impostor equalities). Similarly, the probability for an impostor to find an order $m$ hash chain following Eq. (2) in $\mathcal{U}$ is

$$1 - (1 - P_{fa})^N.$$

$\square$

When using the prefix algorithm described in Table 1, or its randomized variant, the overall size of a biometric negative database is given by the following lemma.

**Lemma 1** *The size of an $(\mathcal{H}, m, \mathcal{DB})$-biometric negative database $\overline{\mathcal{DB}}$ obtained through the prefix algorithm is at most*

$$2 \times N \times \binom{L}{m} \times l^2$$

*where $l$ is the length of the binary representation of an order $m$ hash chain as in Eq. (2). Via its randomized variant, the upper size becomes*

$$2 \times N \times \binom{L}{m} \times l^3.$$

*Proof.* $\overline{\mathcal{DB}}$ is a negative database of $\mathcal{U} = \bigcup_{k=1}^{N} \mathcal{DB}_{b_k}$. Each $\mathcal{DB}_{b_k}$ contains all order $m$ hash chains obtained for the template $b_k$, i.e. all the $m$ choices of hash functions among the $L$ from $\mathcal{H}$. The remaining is deduced from the expansion when the algorithms for negative representation of Section 3 are applied. $\square$

When the original database contains $N$ templates of length $n$, in the deterministic case, the expansion factor is $2 \times \binom{L}{m} \times l^2/n$ and respectively $2 \times \binom{L}{m} \times l^3/n$ in the randomized case.

Note that although we focus here on the expansion when applying the prefix algorithms, our concept is compatible with any algorithm for negative representation generation.

## 4.2 Security Discussion

Concerning the confidentiality of the original templates, the choice of an adequate negative database creation algorithm leads to:

**Proposition 3** *Let $\overline{\mathcal{DB}}$ be an $(\mathcal{H}, m, \mathcal{DB})$-biometric negative database generated through an algorithm which outputs hard instance for the reconstruction problem (cf. Section 3).*

*The knowledge of $\overline{\mathcal{DB}}$ does not allow to retrieve the list of original templates in $\mathcal{DB}$.*

This is straightforward, as by the above assumption, reconstructing $\mathcal{U}$ is hard. Note that even without that reconstructing the original templates might be a difficult task: if one succeeds in retrieving part of the hash chains of $\mathcal{U}$ from $\overline{\mathcal{DB}}$, he still has no idea of which hash chains are related to the same original template. Nevertheless, one additional advantage on using hard instances, as already mentioned in Section 3, is the possibility to create several, and unlinkable, randomized representations. This means that it is possible to have different biometric negative databases from the same authorization list. For instance this fits well to access control use cases where the local terminals can possess the same authorization list, without – thanks to our technique – letting the possibility to make a direct correlation between them.

Other advantages are implied by these negative representations and their randomization, it hides the number of data which are negatively represented and after insertion or deletion of data, applying randomization on a negative database enables to hide the size variation of the original dataset.

## 4.3 Operations

On biometric negative databases as defined in Definition 3, we can apply all the operations available for classical negative databases – see Section 3 – such as the randomization, insertion, deletion, clean-up, relational algebra, ..., operations. Nevertheless all operations on $\overline{\mathcal{DB}}$ have not the same impact with respect to the original database $\mathcal{DB}$.

In particular, we are interested in the two following operations.

Enrollment of a new user, via a capture $b_{N+1}$, is straightforward using the insertion functionality of the negative database for all the hash chains computed from $b_{N+1}$.

Revocation of a user is less easy. Due to the property of the biometric negative database structure, only authorization checks are possible and there is no way to link different hash chains together. So when using deletion functionality of negative databases, with respect to a fresh capture $b'$ which is prior determined as authorized (i.e. close to a $b_k$ for some $k$), you can only delete the hash chains found for $b'$ (i.e. to add in $\overline{\mathcal{DB}}$ the hash chains computed from $b'$). Thus this will not suppress all the hash chains related to $b_k$. Moreover, if the prior authorization corresponds to a False 'Member' Decision, then this revocation would affect other users. This last point is a common problem for anonymous membership checks in biometric systems. Concerning the former issue, we suggest to mitigate it by avoiding deletion of database through $\overline{\mathcal{DB}}$ but by adding a dedicated revocation database (blacklist), which can be represented in a negative form.

### 4.4 Variant for Authentication

The Definition 3 leads to the construction of a biometric negative database for authorization checks purpose. When authentication (1-to-1) is aimed, the variant below is given.

**Definition 4** *An $(\mathcal{H}, m, \mathcal{DB})$-biometric negative database for authentication $\overline{\mathcal{DB}}_{auth}$ is the union of negative representations $\overline{\mathcal{DB}}_{b_k}$ of the datasets $\mathcal{DB}_{b_k}$ $(k = 1 \ldots N)$. More precisely, we defined $\overline{\mathcal{DB}}_{auth}$ as the database containing for all $k$, the elements of $\overline{\mathcal{DB}}_{b_k}$ with $k$ appended:*

$$\overline{\mathcal{DB}}_{auth} = \bigcup_{k=1}^{N} [\overline{\mathcal{DB}}_{b_k}, k].$$

Appending $k$ enables to restrict the verification to one sole $\overline{\mathcal{DB}}_{b_k}$. The associated algorithm for authentication check, with a fresh capture $b'$ and an identity claim $i$, is described in Table 3. Note that this construction can be used as well for identification use cases where no identity claim is provided by running comparisons with all elements of $\overline{\mathcal{DB}}_{auth}$: the list of possible identities would be the list of index $k$ for which $b'$ is not detected as member of $\overline{\mathcal{DB}}_{b_k}$.

---

**Input:** fresh biometric template $b'$ and identity claim $i$

1. result← OK
2. For all $1 \leq j_1 < \ldots < j_m \leq L$
3.     $z \leftarrow j_1 || \ldots || j_m || h_{j_1}(b') || \ldots || h_{j_m}(b')$
4.     If $\mathsf{IsMember}_{\overline{\mathcal{DB}}_{b_i}}(z) = $ OK
5.       Then
6.         result← NOK
7.         Break
8.       Else Continue
9. End For
10. Output result

---

Table 3: Authentication Check for Biometric Negative Database

Similar to the results explained in Section 4.1, we have:

**Lemma 2** *Let $\overline{\mathcal{DB}}_{auth}$ be an $(\mathcal{H}, m, \mathcal{DB})$-biometric negative database for authentication.*

- *The False Reject Rate of the authentication check (Table 3) is $P_{fr}$.*
- *The False Accept Rate of the authentication check (Table 3) is $P_{fa}$.*
- *When using the randomized variant of the prefix algorithm, the size of $\overline{\mathcal{DB}}_{auth}$ is*
$$2 \times N \times \binom{L}{m} \times l^3.$$

Note that here, we give only the size with respect to the randomized variant of the prefix algorithm. We cannot use the deterministic variant as there is no mixing of the different datasets $\mathcal{DB}_k$ that would ensure unlinkability of the hash chains.

**Remark 1** *This is interesting that the expansion factor in the authorization scenario is lower or equal to the expansion factor in the authentication scenario.*

## 5 An Example

We describe now the application of our concept to the example of functions used in [16] for iris identification. The functions used by [16] are $L = 128$ different restrictions of the 2048 bits iris vectors to 10 bits.

Let $\lambda_{min} = 0.25 \cdot 2048 = 512$, $\lambda_{max} = 0.35 \cdot 2048 = 716.8$ be the values of Condition 1. Then the above family of functions is an $(\lambda_{min}, \lambda_{max}, p_1, p_2)$-LSH family, with the probability $p_1$ to obtain the same small proportion for two iriscodes coming from the same user $p_1 = (1 - \frac{\lambda_{min}}{2048})^{10} \simeq 0.056$ and the probability to obtain the same value for two iriscodes coming from different users $p_2 = (1 - \frac{\lambda_{max}}{2048})^{10} \simeq 0.013$.

If we choose a hash chain order $m$ equal to 4, then the binary length of the hash chains is $l = (7 + 10) \times 4 = 68$ and the error rates are $P_{fr} \simeq 0.066$ and $P_{fa} \simeq 0.095$. This shows the interest of using hash chains. This choice of parameteres is valuable for authentication use case. This needs still to be improved to be efficient for authorization use case with medium scale databases.

According to Lemma 1, the upper bound for expansion factors when using the deterministic prefix algorithm will be about $2^{25.5}$ and about $2^{31.6}$ with the randomized prefix algorithm.

As an even more practical example, if we take the order $m$ equal to 3 – which is the threshold used in [16] for determination of candidates in an iris identification scenario – and $N = 100$ enrolled iriscodes from different users, then it leads to an overall biometric negative database of about 20 Go.

## 6 Conclusion

This paper introduces the notion of negative database for biometric data. While the concept of negative database has been introduced in 2004, our proposal is the first one, as far as we know on this very subject. Although the storage of biometric data seems a natural field of application for this concept of negative database, one should understand that what make our work possible are the prior researches on the quantization of biometric. Indeed, our contribution exploits the simpler matching algorithm that this quantization permits. This is not the first time that this new matching algorithms raise results. For instance, beyond their own interest as an alternative to traditional matching algorithms, they are currently considered as a part of the solution for biometric identification in a encrypted way [4].

We believe that our scheme can still be optimized. For example, we directly store in the positive database $\mathcal{DB}$ the biometric data taken during the enrollment phase. Whereas, we are interested in the whole Hamming ball of radius $\lambda_{min}$. A clever representation of biometric data should lead to more compact positive database and negative database.